\documentclass[%
 reprint,
showpacs,
 amsmath,amssymb,
prb,
]{revtex4-1}
\usepackage{graphicx}
\usepackage{dcolumn}
\usepackage{bm}
\usepackage{braket}
\usepackage{color}
\usepackage{accents}
\usepackage{enumitem}

\begin{document}
\title{
Odd-parity pairing correlations in a $d$-wave superconductor
}
\author{Jaechul Lee$^{1}$}
\author{Satoshi Ikegaya$^{2}$}
\author{Yasuhiro Asano$^{1,3}$}%
\affiliation{
$^{1}$Department of Applied Physics, Hokkaido University, Sapporo 060-8628, Japan\\
$^{2}$ Max-Planck-Institut f\"{u}r Festk\"{o}rperforschung,
Heisenbergstrasse 1, D-70569 Stuttgart, Germany\\
$^{3}$Center of Topological Science and Technology,
Hokkaido University, Sapporo 060-8628, Japan
}
\date{\today}
\begin{abstract}
We theoretically study the effects of spin-orbit interactions on 
symmetry of a Cooper pair in a spin-singlet $d$-wave superconductor 
and those on the chiral property of the surface Andreev bound states at the zero energy.
The pairing symmetry is analyzed by using the anomalous Green's function 
which is obtained by solving the Gor'kov equation analytically. 
The chiral property of surface bounds states is discussed by using an index 
which represents a number of the zero-energy states in the presence of potential disorder at a surface.
A spin-orbit interaction induces a spatially uniform spin-triplet $p$-wave pairing correlation 
in a superconductor and an odd-frequency spin-triplet $s$-wave pairing correlation at a surface.
The spin-orbit interaction splits the Fermi surface into two depending on spin configuration.
As a result of splitting, the index can be nontrivial nonzero values.
On the basis of the close relationship among the odd-frequency $s$-wave pairing correlation at a surface, 
the nonzero index of surface bound states and the anomalous proximity effect,
we provide a design of a superconductor which causes the strong anomalous proximity effect.
\end{abstract}
\maketitle

\section{Introduction}
Proximity structures consisting of a dirty normal metal and spin-triplet superconductors 
indicate unusual electric transport properties such as 
the quantization of zero-bias conductance in a normal-metal/superconductor
(NS) junction and the fractional current-phase relationship of the Josephson current in a
superconductor/normal-metal/superconductor (SNS) junction.~\cite{tanaka:prb2004,asano:prl2006} 
Such unusual transport phenomena via a dirty normal metal are called anomalous proximity effect.
Topologically protected bound states at a surface of a spin-triplet superconductor 
play a key role in the anomalous proximity effect.
Namely, they penetrate into a dirty normal metal and form the resonant transmission channels 
at the Fermi level, which causes the perfect electron transmission through the dirty 
normal metal~\cite{ikegaya:prb2015,ikegaya:jphys2016,ikegaya:prb2016}.
The formation of the resonant states at the Fermi level can be detected directly as
a large zero-energy peak in the local density of states (LDOS)~\cite{tanaka:prb2004,asano:prl2006} 
at a dirty normal metal.
The anomalous proximity effect has been considered as a part of Majorana physics
~\cite{sato:prb2006,sato:prb2009,lutchyn:prl2010,oreg:prl2010,asano:prb2013} 
because a spin-triplet superconductor hosts Majorana fermions at its surface.
Unfortunately, well-established spin-triplet superconductors have never been 
discovered yet.  

Tamura and Tanaka~\cite{tamura:prb2019} have studied theoretically
the LDOS at a dirty normal metal attached 
to a $d_{xy}$-wave superconducting film with the Rashba spin-orbit interaction, where 
the NS interface is parallel to the $y$ direction as illustrated in Figs.~\ref{fig:system}(a)-(c).
Their numerical results indicate signs of the anomalous proximity effect.
The modest enhancement of the LDOS at the zero energy suggests
the penetration of the zero-energy states (ZESs) into a dirty metal.
In addition, they found an odd-frequency spin-triplet $s$-wave Cooper pair in the normal 
metal. (See recent review papers on odd-frequency pairing correlations.\cite{linder:rmp2019,cayao:epj2020}) 
Although the signal of the proximity effect is very weak, the 
results are highly nontrivial due to reasons as follows.
 It has been already established in the absence of spin-orbit interactions that 
a $d_{xy}$-wave superconductor in Fig.~\ref{fig:system} (c) does not exhibit
any proximity effect in a dirty metal~\cite{asano:prb2001-dwave,asano:jpsj2002,tanaka:prl2003}. 
Spin-orbit interaction may induce a spin-triplet Cooper pair in a spin-singlet 
superconductor~\cite{bauer:prl2004,reyren:science2007,gorkov:prl2001,frigeri:prl2004,fujimoto:jpsj2007,
sato:prb2009,reeg:prb2015,bobkova:prb2017,cayao:prb2018}. 
However, mechanisms of the symmetry conversion to an odd-frequency $s$-wave Cooper are still unclear. 


In this paper, we theoretically study the effects of spin-orbit interactions in a spin-singlet 
$d_{xy}$-wave superconductor on symmetry of a Cooper pair and on chiral property of Andreev bound states at
its surface.
We analyze the symmetry of a Cooper pair by using the anomalous Green's function obtained by solving 
the Gor'kov equation analytically.
The results show that a specific spin-orbit interaction generates 
a spin-triplet $p_x$-wave pairing correlation in bulk and  
an odd-frequency spin-triplet $s$-wave pairing correlation at a surface.
The Bogoliubov-de Gennes (BdG) Hamiltonian of such a $d_{xy}$-wave superconductor preserves a nontrivial 
chiral symmetry, which enable us to define an index by using the chiral eigenvalues of 
surface Andreev bound states at the zero energy. 
We find that the index, a measure of the strength of the anomalous proximity effect, can be
nonzero values only in the presence of the spin-orbit interaction.
On the basis of obtained results, we explain why 
the signs of the anomalous proximity effect in Ref.~\onlinecite{tamura:prb2019} is weak 
and provide a design for a superconductor which causes the strong anomalous proximity effect.

This paper is organized as follows. 
In Sec.~II, we explain the anomalous proximity effect in more detail and three necessary conditions 
of the BdG Hamiltonian for the anomalous proximity effect, which makes more clear a goal of this paper.
In Sec.~III, we analyze the chiral property of the surface bound states at the zero energy. 
In Sec.~IV, symmetry of a Cooper pair appearing at a surface of 
a $d_{xy}$-wave superconductor with spin-orbit interactions are analyzed by using  
the anomalous Green's function.
In Sec.~V, we discuss the anomalous proximity effect when the two types of spin-orbit interaction 
coexist.
The conclusion is given in Sec.~VI.
In the text of this paper, we use the system of units $\hbar=k_B=c=1$, where $k_B$ is the Boltzmann constant 
and $c$ is speed of light.

\section{Anomalous proximity effect}
\label{sec:ape}
In this section, we explain the conductance quantization in an NS junction 
due to anomalous proximity effect, 
necessary conditions for a superconductor exhibiting the anomalous proximity effect, 
and goals of this paper.
 
\subsection{chiral property of ZESs}
\label{ssec:chiral}

The most drastic phenomenon of the anomalous proximity effect may be the quantization of the 
differential conductance in a NS junction shown in Fig.~\ref{fig:system}(c). 
The conductance at the zero bias can be described by
 \begin{align}
 G_{\mathrm{NS}}=\frac{2e^2}{h}|\mathcal{N}_{\textrm{ZES}}|,\label{eq:gnsd}
 \end{align}
in the limit of strong potential disorder in a normal metal, 
where $|\mathcal{N}_{\textrm{ZES}}|$ represents a number of the zero-energy states (ZESs) 
remaining
at a dirty surface of a superconductor.~\cite{tanaka:prb2004,ikegaya:prb2016}
We first explain properties of the index $\mathcal{N}_{\textrm{ZES}}$ in 
the case of a spin-triplet $p_x$-wave superconductor which is an example of 
superconductor exhibiting the anomalous proximity effect.
At a clean surface of a $p_x$-wave superconductor, topologically protected bound states 
appear at the zero energy as a result of a nontrivial winding number
 $\mathcal{W}(k_y)=\pm 1$ in one-dimensional Brillouin zone.~\cite{sato:prb2011} 
In Fig.~\ref{fig:sabs}(a) in Appendix~\ref{asec:sabs}, we plot the 
eigenvalues of the BdG Hamiltonian for a $p_x$-wave superconductor 
on a two-dimensional tight-binding lattice.
The superconducting gap has two nodes at $k_y = \pm k_F$ with $k_F= \pi/2$ being the Fermi wavenumber.
The degree of the degeneracy at the zero energy is equal to the number of propagating channels on the 
Fermi surface $N_c$ because each propagating channel $-k_F < k_y < k_F$ 
accommodates a ZES at a surface.
We also show the results for a $d_{xy}$-wave superconductor in Fig.~\ref{fig:sabs}(b). 
The winding number $\mathcal{W}(k_y)$ can be defined only in the 
presence of the translational symmetry in the $y$ direction. 
Therefore, the large degree of degeneracy at the zero energy in both Figs.~\ref{fig:sabs}(a) and (b) is a 
direct consequence of translational symmetry of the Hamiltonian in the $y$ direction. 
The zero-bias conductance in such a ballistic NS junction is described by
 \begin{align}
 G_{\mathrm{NS}}=\frac{2e^2}{h} {N}_{c}.
 \end{align}
The results are valid for both a $p_x$-wave junction and a $d_{xy}$-wave junction.

Random impurities near the surface may lift the degeneracy at the zero energy 
because they break translational symmetry.
The index $|\mathcal{N}_{\textrm{ZES}}|$ in Eq.~(\ref{eq:gnsd}) represents the number of the ZESs 
remains even in the presence of potential disorder. 
As discussed in detail later, the index $\mathcal{N}_{\textrm{ZES}}$ can be defined when the 
Bogoliubov-de Gennes Hamiltonian preserves a chiral symmetry.
Therefore we discuss fundamental symmetries of BdG Hamiltonian in this paragraph.
The BdG Hamiltonian in momentum space is represented as
\begin{align}
\check{H}_{\mathrm{BdG}}(\boldsymbol{k})
=&\left[ 
\begin{array}{cc}
\hat{H}_{\mathrm{N}}(\boldsymbol{k}) & \hat{\Delta}(\boldsymbol{k}) \\
-\hat{\Delta}^\ast(-\boldsymbol{k}) & - {\hat{H}}_{\mathrm{N}}^\ast(-\boldsymbol{k})
\end{array}
\right], \label{eq:hbdg}
\end{align}
where $\hat{H}_{\mathrm{N}}(\boldsymbol{k})$ and $\hat{\Delta}(\boldsymbol{k})$ 
are the normal state Hamiltonian and the pair potential, respectively.
Throughput this paper, symbols $\check{\cdots}$ and $\hat{\cdots}$ represent 
$4 \times 4$ and $2 \times 2$ matrices, respectively.
The Hamiltonian for a $p_x$-wave superconductor in Eq.~(\ref{aeq:hpx}) preserves time-reversal symmetry,
\begin{align}
\mathcal{T}_-\, \check{H}_{\mathrm{BdG}}\, \mathcal{T}_-^{-1}= \check{H}_{\mathrm{BdG}}, \quad 
\mathcal{T}_-= i\hat{\sigma}_2\,  \mathcal{K},
\end{align}
where $\mathcal{K}$ denotes taking the complex conjugation for Hamiltonian in real space, 
and taking the complex conjugation plus applying $\boldsymbol{k}\to - \boldsymbol{k}$ 
for Hamiltonian in momentum space.
The Pauli's matrix in spin space is denoted by $\hat{\sigma}_j$ for $j=1-3$. 
A BdG Hamiltonian always preserves particle-hole symmetry, 
\begin{align}
\mathcal{C}\, \check{H}_{\mathrm{BdG}}\, \mathcal{C}^{-1}= -\check{H}_{\mathrm{BdG}}, \quad 
\mathcal{C}=  \hat{\tau}_1\, \mathcal{K},
\end{align}
where $\hat{\tau}_j$ for $j=1-3$ is the Pauli's matrix in particle-hole space.
By combining the two symmetries, any Hamiltonian belonging to class DIII preserves a chiral symmetry
\begin{align}
\left\{ \check{\Gamma}_{\mathrm{DIII}}, \, \check{H}_{\mathrm{BdG}} \right\}=0, \quad
\check{\Gamma}_{\mathrm{DIII}}= \hat{\tau}_1\, \hat{\sigma}_2. \label{eq:d3chiral}
\end{align}
Since $\check{\Gamma}^2_{\mathrm{DIII}}=1$, the eigenvalue of the chiral operator is either 1 or $-1$.
A ZES is always an eigenstate the chiral operator~\cite{sato:prb2011}. 
Indeed, it is easy to confirm that the Hamiltonian in 
Eq.~(\ref{aeq:hpx}) anticommutes to 
$\check{\Gamma}_{\mathrm{DIII}}$ and that the wave function of surface bound states at the zero energy
 $\phi_{p_x, \pm}$ in Eq.~(\ref{aeq:phi_pm}) is the eigenfunction of 
$\check{\Gamma}_{\mathrm{DIII}}$ belonging to the chial eigenvalue of $\pm 1$.
The index in Eq.~(\ref{eq:gnsd}) can be defined by
\begin{align}
\mathcal{N}_{\textrm{ZES}} \equiv N_+ - N_-, \label{eq:nzes_def}
\end{align}
where $N_\pm$ is the number of the ZESs belonging to the chiral eigenvalue of $\pm 1$.
The index is an invariant as long as the Hamiltonian preserves the chiral symmetry  in Eq.~(\ref{eq:d3chiral}).
Thus $\mathcal{N}_{\textrm{ZES}}$ calculated in a clean superconductor remains unchanged 
even in the presence of potential disorder because
the random impurity potential $V_{\mathrm{imp}}(\boldsymbol{r})\, \hat{\tau}_3$ preservers the chiral 
symmetry in Eq.~(\ref{eq:d3chiral}).
The anomalous proximity effect happens when the index takes a nontrivial value $\mathcal{N}_{\textrm{ZES}}\neq 0$.
However, ZESs of $H_{\mathrm{BdG}}$ in class DIII always satisfy $\mathcal{N}_{\textrm{ZES}}= 0$ 
in terms of their chiral eigenvalues of $\check{\Gamma}_{\mathrm{DIII}}$~\cite{ikegaya:prb2018}. 
Therefore, the BdG Hamiltonian of a superconductor exhibiting the anomalous proximity effect must satisfy
following necessary conditions,
\begin{enumerate}[label=(\roman*)]
\item $H_{\mathrm{BdG}}$ anticommutes to an extra chiral operator other than $\check{\Gamma}_{\mathrm{DIII}}$,
\item Zero-energy states at a surface of a superconductor have such a chiral property as 
$\mathcal{N}_{\textrm{ZES}}\neq 0$ in terms of the extra chiral operator.
\end{enumerate}
We show a simple way to find an extra chiral operator as follows.
The Hamiltonian in Eq.~(\ref{aeq:hpx}) remains unchanged under the rotation 
around the first axis in spin space,
\begin{align}
\check{R}_1\, \check{H}_{p_x}\, \check{R}_1^{-1}=  \check{H}, \quad
\check{R}_1= \hat{\sigma}_1\, \hat{\tau}_3.
\end{align}
Using such a spin rotation symmetry it is possible to define a time-reversal like
symmetry of Hamiltonian
\begin{align}
\mathcal{T}_+\, \check{H}_{p_x} \, \mathcal{T}_+^{-1}= \check{H}_{p_x}, \quad 
\mathcal{T}_+= \check{R}_1 \mathcal{T}_- = - \hat{\sigma}_3\, \hat{\tau}_3 \, \mathcal{K}.
\end{align}
As a result of the spin-rotation symmetry of the Hamiltonian, we find
\begin{align}
\{\check{\Gamma}_{\mathrm{BDI}}, \, \check{H}_{p_x} \}=0, \quad 
\check{\Gamma}_{\mathrm{BDI}}= i \mathcal{T}_+ \, \mathcal{C} = \hat{\sigma}_3\, \hat{\tau}_2.
\label{eq:bdi_chiral}
\end{align}
Since $\mathcal{T}_+^2=1$, $\check{H}_{p_x}$ belongs also to class BDI.
The wave function of ZESs at a surface of a $p_x$-wave superconductor 
can be represented alternatively as $\phi_{p_x, \sigma}$ in Eq.~(\ref{aeq:phi_px}).
We find that $\phi_{p_x, \sigma}$ belongs to the positive chiral eigenvalues 
of $\check{\Gamma}_{\mathrm{BDI}}$ irrespective of $\sigma$. 
Therefore we find $N_+=N_c$ and $N_-=0$, which results in $\mathcal{N}_{\mathrm{ZES}}=N_c$.
A $p_x$-wave superconductor satisfies the conditions (i) and (ii) in terms of
the chiral operator $\check{\Gamma}_{\mathrm{BDI}}$.
Two of authors showed that some superconductors under strong Zeeman field in the presence of spin-orbit 
interactions preserve such an extra chiral symmetry
and host flat-band ZESs at their dirty surfaces.~\cite{ikegaya:prb2018} 
The anomalous proximity effect has been considered to be a part of Majorana physics because 
such artificial superconductors host Majorana fermions at their surface.~\cite{alicea:prb2010,you:prb2013}
At present, however, we are thinking that a superconductor hosting Majorana fermions 
is a sufficient condition for the anomalous proximity effect.
The necessary conditions (i) and (ii) are more relaxed than those generating Majorana fermions.

\subsection{An odd-frequency $s$-wave pair}
The anomalous proximity effect has two aspects:
the penetration of a quasiparticle at the zero energy into a dirty metal as 
discussed in Sec.~\ref{ssec:chiral}
and the penetration of an odd-frequency Cooper pair into a dirty metal.  
The existence of $\mathcal{N}_{\mathrm{ZES}}$-fold degenerate ZESs 
can be observed as a large zero-energy peak in the 
local density of states (LDOS) at a dirty metal. 
In the mean-field theory of superconductivity, 
the density of states can be calculated from the normal Green's function $\hat{G}$ 
and the pairing correlations are described by the anomalous Green's function $\hat{F}$.
As the two Green's functions are related to each other through the Gor'kov equation, 
the enhancement of $\hat{G}$ at the zero energy causes an anomaly of $\hat{F}$ at the zero energy. 
The anomalous Green's function in Matsubara
representation $\hat{F}( \boldsymbol{k}, i\omega_n)$ enables us to analyze symmetry of 
pairing correlations, where $\omega_n=(2n+1)\pi T$ is a fermionic Matsubara frequency 
and $T$ is a temperature.
The anomalous Green's function obeys the antisymmetric relation under the permutation of two electrons 
consisting of a Cooper pair,
\begin{align}
\hat{F}^{\mathrm{T}}(- \boldsymbol{k}, -i\omega_n)=&- \hat{F}( \boldsymbol{k}, i\omega_n), \label{eq:antisymm_f}
\end{align}
where $\mathrm{T}$ means the transpose of a matrix.
In the standard representation, the anomalous Green's function is decomposed into four spin components as
\begin{align}
\hat{F}( \boldsymbol{k}, i\omega_n)=&i\left[f_0(\boldsymbol{k}, i\omega_n)+ \boldsymbol{f}(\boldsymbol{k}, i\omega_n)\cdot \hat{\boldsymbol{\sigma}} \right] \, \hat{\sigma}_2, \label{eq:def_deco}
\end{align}
where $f_0$ is a spin-singlet component and $\boldsymbol{f}$ represents three spin-triplet 
components. In a dirty normal, the diffusive motion of a quasiparticle makes 
the Green's function be isotropic in momentum space as 
\begin{align}
\hat{F}( i\omega_n)=&i\left[f_0(i\omega_n)+ \boldsymbol{f}(i\omega_n)\cdot \hat{\boldsymbol{\sigma}} \right] \, \hat{\sigma}_2.
\end{align}
The relation in Eq.~(\ref{eq:antisymm_f}) implies that the spin-singlet component is an even function of $\omega_n$ 
as $f_0(-i\omega_n)=f_0(i\omega_n)$.
The spin-triplet components, on the other hand, are odd function of $\omega_n$ as
 $\boldsymbol{f}(-i\omega_n)=-\boldsymbol{f}(i\omega_n)$. 
To compensate the anomaly of $\hat{G}$ at $\omega_n \to 0$, the anomalous Green's 
function must have a component of $\hat{F}(i\omega_n) \propto 1/\omega_n$.~\cite{tanaka:prl2007}
Such an odd-frequency pair must be spin-triplet.
Thus the last necessary condition for the anomalous proximity effect is
\begin{enumerate}[label=(\roman*)]
\setcounter{enumi}{2}
\item An odd-frequency spin-triplet $s$-wave pairing correlation $\boldsymbol{f} \propto 1/\omega_n$ 
exists at a surface as a result of the large zero-energy peak in the local density of states.
\end{enumerate}
In a spin-singlet $d_{xy}$-wave superconductor, however,
no spin-triplet pairing correlation exists in the absence of spin-dependent potentials.

\subsection{Goals of this paper}
To make clear a motivation of this study, we show that a spin-singlet $d_{xy}$-wave superconductor 
without spin-orbit interaction does not cause the anomalous proximity effect.
A spin-singlet $d_{xy}$-wave pair potential is defined by,
\begin{align}
\hat{\Delta}(\boldsymbol{k}) = \Delta_{\boldsymbol{k}}\, i \hat{\sigma}_2, 
\quad \Delta_{\boldsymbol{k}} = \Delta \, \bar{k}_x \bar{k}_y,
\end{align}
where $k_x (k_y)$ is the wavenumber in the $x (y)$ direction. 
The wave numbers in the pair potential are normalized to the Fermi wavenumber $k_F$ as
$\bar{k}_x=k_x / k_F$ and $\bar{k}_y=k_y / k_F$.
It is easy to confirm that the Hamiltonian for a $d_{xy}$-wave superconductor in 
Eq.~(\ref{aeq:hdxy}) anticommutes to $\check{\Gamma}_{\mathrm{BDI}}$ and that
the wave function of the ZESs at a surface of a $d_{xy}$-wave superconductor are given in 
Eq.~(\ref{aeq:phi_dxy}). 
Their chiral eigenvalues depend on $\mathrm{sgn}(k_y)$  
reflecting the sign change of the pair potential at $k_y=0$.
As displayed in Fig.~\ref{fig:sabs}(b), the number of the ZESs at $k_y>0$ is equal to that at $k_y<0$. 
Although Eq.~(\ref{aeq:hdxy}) satisfy condition (i), 
the index $\mathcal{N}_{\mathrm{ZES}}$ is always zero in the absence of spin-orbit interactions.
Therefore, we concluded that random impurity potential completely lifts the high degeneracy at 
the zero energy in Fig.~\ref{fig:sabs}(b) and that a $d_{xy}$-wave superconductor does not 
show the proximity effect in the absence of spin-dependent potentials.

In the presence of Rashba spin-orbit interaction in a $d_{xy}$-wave superconductor, 
numerical results of LDOS and those of the pairing correlation in a dirty metal 
indicate that condition (iii) may be satisfied~\cite{tamura:prb2019}.  
The BdG Hamiltonian in momentum space reads,
\begin{align}
\check{H}_{0}
 =& (\xi_{\boldsymbol{k}} -\lambda k_x \hat{\sigma}_2) \, \hat{\tau}_3 +\lambda k_y \hat{\sigma}_1 
- \Delta_{{\boldsymbol{k}}} \hat{\tau}_2\, \hat{\sigma}_2
\label{eq:h0}, 
\end{align}
where $\xi_{\boldsymbol{k}}=\boldsymbol{k}^2/{2m}-\epsilon_F$ is the kinetic energy of a quasiparticle measured from the Fermi 
energy $\epsilon_F= k_F^2/(2m)$ and $\lambda$ represents the strength of the spin-orbit interaction.
When the two spin-orbit interaction terms 
 $-\lambda k_x \hat{\sigma}_2 \, \hat{\tau}_3 $ and $\lambda k_y \hat{\sigma}_1$ coexist,  
$\check{\Gamma}_{\mathrm{DIII}}$ is the only chiral operator anticommutes to $\check{H}_{0}$.
Since $\mathcal{N}_{\mathrm{ZES}}=0$ in terms of the chiral eigenvalues of $\check{\Gamma}_{\mathrm{DIII}}$,
the degeneracy of the ZESs is fragile in the presence of potential disorder.~\cite{ikegaya:prb2018}
Thus we infer that the anomalous proximity effect in such junction would be very weak.
Namely, the zero-bias conductance in an NS junction would not be quantized and 
the zero-energy peak in LDOS would disappear in the limit of strong potential disorder. 

In such a situation, the goal of this paper is to make clear roles of the spin-orbit interaction 
terms in the anomalous proximity effect.
To achieve the objective, we analyze the chiral property of the surface bound states of two 
different superconductors: the Hamiltonian of one superconductor contains only a spin-orbit interaction $\lambda k_y \hat{\sigma}_1$
and that of the other contains only $-\lambda k_x \hat{\sigma}_2 \, \hat{\tau}_3$.
After showing if $\mathcal{N}_{\mathrm{ZES}}$ is 0 or not for the two superconductors,  
we make clear how an odd-frequency spin-triplet $s$-wave pair appears at a surface.
On the basis of the obtained results, we provide a theoretical design of a superconductor that 
exhibits the strong anomalous proximity effect.

\section{Two types of superconductor}
\label{sec:surface_state}

\begin{figure}[tbp]
\begin{center}
  \includegraphics[width=7.0cm]{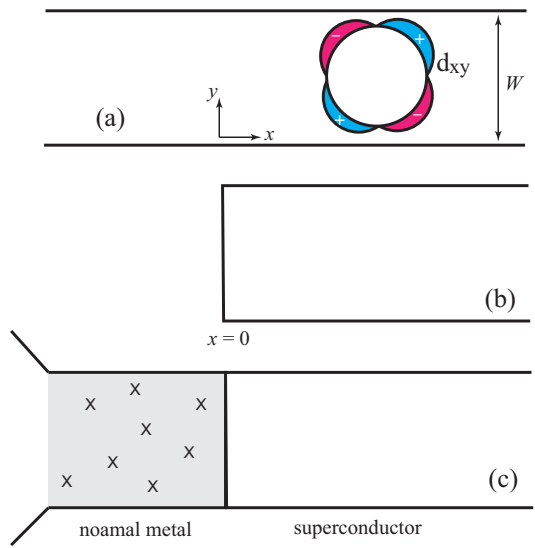}
\caption{Schematic pictures of superconducting proximity structures.
A spin-singlet $d_{xy}$-wave superconductor is infinitely long in the $x$ direction in (a).
The width of the superconductor is $W$ in the $y$ direction. 
The Green's function at a surface is calculated for a semi-infinite superconductor 
as shown in (b).
We attach a normal metal to a superconductor at $x=0$ in (c), 
where "cross" symbols represent impurities.
A Cooper pair penetrates into the normal metal and causes the proximity
effect.
 }
\label{fig:system}
\end{center}
\end{figure}

We divide the Rashba spin-orbit 
interaction into two parts: $\lambda k_y \hat{\sigma}_1$ 
and $-\lambda k_x \hat{\sigma}_2\, \hat{\tau}_3$ to 
study how they 
modify independently the chiral properties of zero-energy states and the symmetry of a Cooper 
pair at a surface.
Fig.~\ref{fig:system} shows the schematic pictures of a superconductor 
under consideration. A superconductor in Fig.~\ref{fig:system}(a) is infinitely long 
in the $x$ direction 
and the width of the superconductor is $W$ in the $y$ direction. We apply the periodic 
boundary condition in the $y$ direction.
In what follows, we analyze the two BdG Hamiltonians given by
\begin{align}
\check{H}_1=
 \xi_{\boldsymbol{k}} \hat{\tau}_3 +\lambda k_y \hat{\sigma}_1 
- \Delta_{{\boldsymbol{k}}} \hat{\tau}_2\, \hat{\sigma}_2
\label{eq:h1},
\end{align}
and
\begin{align}
\check{H}_2
=&\xi_{\boldsymbol{k}} \hat{\tau}_3 - \lambda k_x \hat{\sigma}_2 \hat{\tau}_3 
- \Delta_{{\boldsymbol{k}}} \hat{\tau}_2\, \hat{\sigma}_2 \label{eq:h2}.
\end{align}
We assume the relation
\begin{align}
\Delta \ll \lambda k_F \ll \epsilon_F,
\end{align}
and discuss the effects of the spin-orbit interaction on the superconducting 
states within the first order of 
\begin{align}
\alpha \equiv \frac{\lambda k_F}{2 \epsilon_F} \ll 1. \label{eq:alpha_def}
\end{align}

The main purpose of the next two subsections are 
checking if the two Hamiltonian preserve extra chiral symmetry and 
calculating the index $\mathcal{N}_{\mathrm{ZES}}$.
In addition, we also define 
the wave number on the Fermi surface and the number of the propagating channels $N_c$,
which are necessary items to represent the anomalous Green's function in Sec.~IV. 

\subsection{ Surface bound states of $\check{H}_1$}
\label{ssec:abs_h1}
\begin{figure}[tbp]
\begin{center}
  \includegraphics[width=8.5cm]{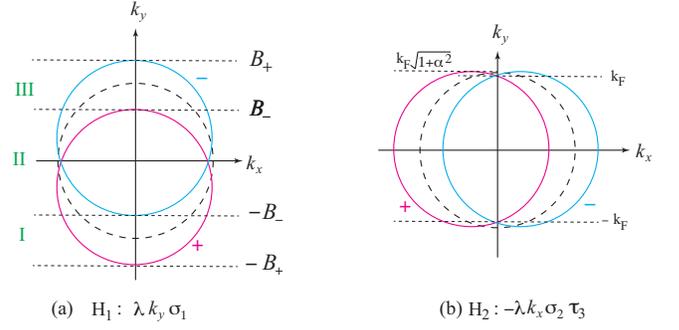}
\caption{
The two Fermi surfaces for $H_1$ and those for $H_2$ 
are illustrated in (a) and (b), respectively.
In (a), the lower (upper) circle indicated by $"+"$ ($"-"$) represents a Fermi surface 
calculated from $\xi_{1,+}(\boldsymbol{k})=0$ ($\xi_{1,-}(\boldsymbol{k})=0$). 
In (b), the left (right) circle represent a Fermi surface 
calculated from $\xi_{2,+}(\boldsymbol{k})=0$ ($\xi_{2,-}(\boldsymbol{k})=0$).  
 }
\label{fig:fermi_surface}
\end{center}
\end{figure}

\begin{figure}[tb]
\begin{center}
  \includegraphics[width=8.5cm]{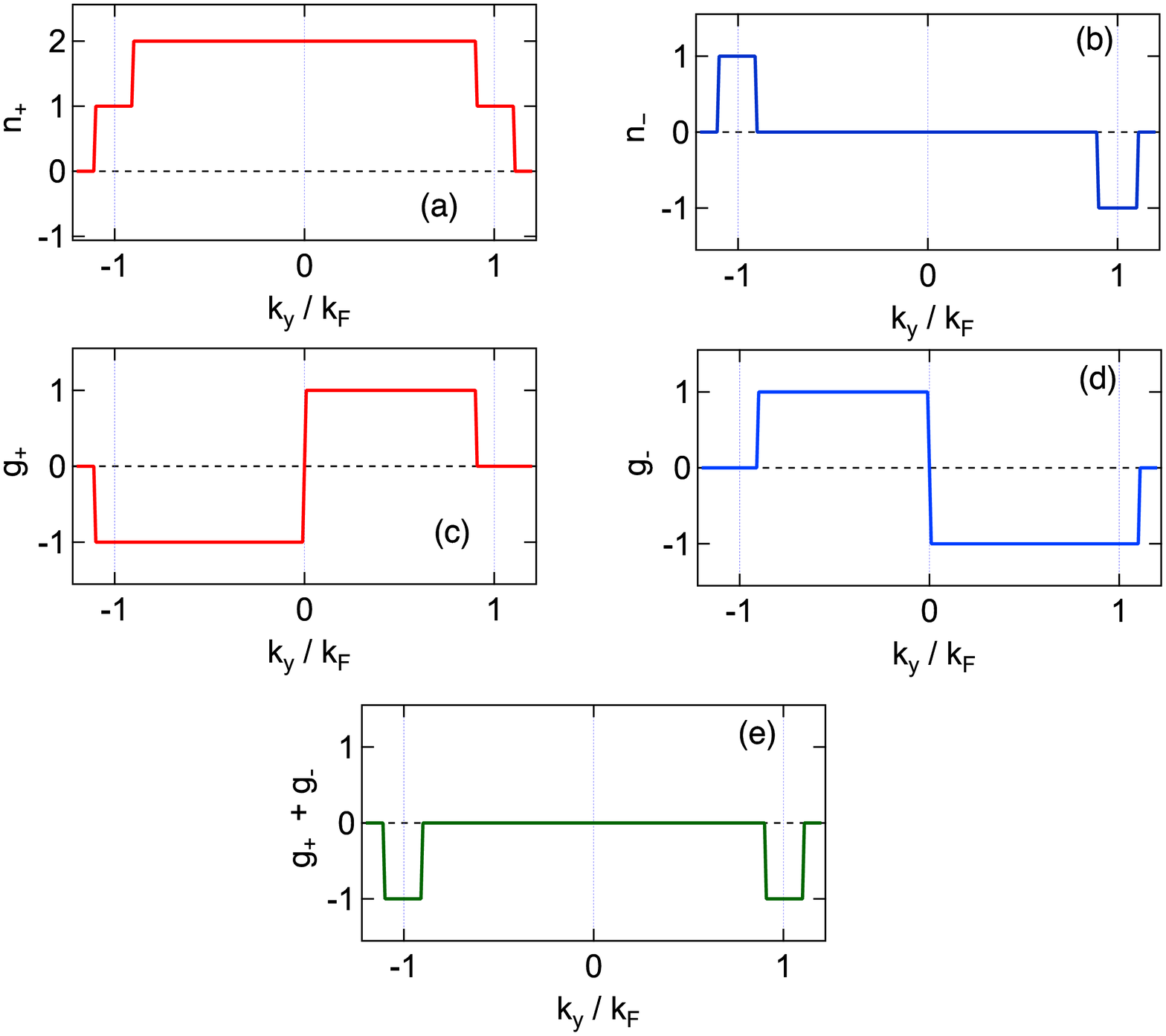}
\caption{In (a), $n_+(k_y)$ represents the number of propagating channels at $k_y$.
 $n_-(k_y)$ in (b) is an odd function of $k_y$. The chiral eigenvalues of 
 the ZESs $g_{+}$ and $g_{-}$ are plotted in (c) and (d), respectively. 
In (e), the summation of the chiral eigenvalues $g_{+}+g_{-}$ is shown.
$n_-(k_y)$ in (b) and $g_{+}+g_{-}$ in (e) are nonzero at (I) and (III).
  }
\label{fig:index}
\end{center}
\end{figure}

We first focus on the $\check{H}_1$ in Eq.~(\ref{eq:h1}).
The positive eigenvalues of $\check{H}_1$ are calculated to be
\begin{align}
E_{1,\pm}= \sqrt{\xi_{1,\pm}^2 +\Delta_{\boldsymbol{k}}^2}, \quad
\xi_{1,\pm}(k, k_y) =& \xi_{\boldsymbol{k}}\pm \lambda\, k_y.
\end{align}
The two Fermi surfaces characterized by $\xi_{1,\pm}=0$ are illustrated in 
Fig.~\ref{fig:fermi_surface}(a).
A wave number in the $y$ direction $k_y$ indicates a transport channel.
As shown in Fig.~\ref{fig:fermi_surface}(a), $k_y$ axis are divided into three 
regions: 
(I) $- B_+  \leq k_y \leq - B_-$, (II) $- B_-  \leq k_y \leq B_-$, and (III) 
 $B_-  \leq k_y \leq B_+$,  
with $B_\pm= [\sqrt{1+\alpha^2} \pm \alpha] k_F$.
The wave numbers
in the $x$ direction are calculated as 
\begin{align}
p_{1\pm} \equiv & \left[k_F^2 - k_y^2 \mp 2 \alpha k_y  k_F \right]^{1/2},
\label{eq:p1_def}
\end{align}
as a function of $k_y$.
A transport channel at $k_y$ on the Fermi surface of $\pm$ branch
is propagating for $p_{1\pm}^2 >0$
and 
is evanescent for $p_{1\pm}^2 <0 $.
Therefore, 
\begin{align}
n_+(k_y)\equiv \Theta(p_{1+}^2)+\Theta(p_{1-}^2),
\end{align}
represents the number of the propagating channels at $k_y$, where $\Theta(x)$ is the step function.
We also define
\begin{align}
n_-(k_y)\equiv \Theta(p_{1+}^2)-\Theta(p_{1-}^2),
\end{align}
for the latter use. 
In Figs.~\ref{fig:index}(a) and (b), we plot $n_\pm$ as a function of $k_y$.
As shown in Fig~\ref{fig:index}(a), both $\pm$ branches are propagating in (II), 
whereas only one branch is propagating in (I) and (III).
The number of propagating channels is calculated as
\begin{align}
N_c\equiv & \sum_{k_y} n_+(k_y) = \frac{W}{2\pi}\int_{-\infty}^\infty dk_y\, n_+(k_y),\\ 
=& \left[ \frac{2 W k_F}{\pi} \right]_{\mathrm{G}}, \label{eq:nc_def}
\end{align}
where $[ \cdots ]_{\mathrm{G}}$ is the Gauss's symbol meaning the integer part of the argument.
Since a propagating channel hosts a ZES, the number of the ZESs at a surface is $N_c$.
A superconductor described by $H_1$
hosts highly degenerate ZESs at its surface around $x \gtrsim 0$ in Fig.~\ref{fig:system}(b). 
As shown in Fig.~\ref{fig:index}(b), $n_-$ is an odd function of $k_y$.
We use this property when we describe an induced pairing correlation by a spin-orbit
interaction in Sec.~\ref{ssec:pc_h1}.

%

The wave functions of a surface bound states at the zero energy 
are described by
\begin{align}
\psi_{1+}(x)=& 
\left[\begin{array}{c} 1 \\ 1 \\ i s_{k} \\ -i s_{k}
\end{array}\right]e^{- q_{x} x} \, \sin p_{1+} x,\label{eq:wf_zes1p} \\
\psi_{1 -}(x)=& 
\left[\begin{array}{c} -1 \\ 1 \\ i s_{k} \\ i s_{k}
\end{array}\right]e^{-q_{x} x} \, \sin p_{1-} x,\label{eq:wf_zes1m}\\
s_{k}=&\textrm{sgn}( k_y ).
\end{align}
Here the wavenumber in superconducting state is given approximately by
\begin{align}
\left[p_{1s}^2  \pm 2m \Delta |\bar{p}_{1s} \bar{k}_y|
 \right]^{1/2}\approx
p_{1s} \pm i  q_{x},
\end{align}
for $s= \pm$.
The imaginary part of the wave number is estimated at $\epsilon=0$ 
as $q_{x} = ( \Delta/ v_F) |\bar{k}_y|$ 
with $ v_F=  k_F/m$ being the Fermi velocity.
As shown in Eqs.~(\ref{eq:wf_zes1p}) and (\ref{eq:wf_zes1m}), 
$q_x$ characterizes the spatial area of the surface bound states. 
The Hamiltonian Eq.~(\ref{eq:h1}) anticommutes to the extra chiral operator
$\check{\Gamma}_{\mathrm{BDI}}$ in Eq.~(\ref{eq:bdi_chiral}).
It is easy to show 
\begin{align}
\check{\Gamma}_{\mathrm{BDI}}\, \psi_{1 \pm}= g_\pm
 \psi_{1 \pm}, \quad g_\pm = \pm s_{k}.
\end{align}
The chiral eigenvalues $g_{+}$, $g_{-}$ and $g_{+}+g_{-}$ are plotted 
as a function of $k_y$ in Figs.~\ref{fig:index}(c), (d) and (e), respectively.
By reflecting the sign change of the pair potential, both $g_{+}$ and $g_{-}$ 
change their signs at $k_y=0$ as shown in Figs.~\ref{fig:index}(c) and (d). 
The index in terms of the chiral eigenvalue of $\check{\Gamma}_{\mathrm{BDI}}$ 
can be calculated as
\begin{align}
\mathcal{N}_{\mathrm{ZES}}=& \sum_{k_y} \left[ g_{+} + g_{-}\right]
=
 \left[ -\frac{2Wk_F}{\pi} \alpha\right]_\mathrm{G}. \label{eq:nzes_bd1}
\end{align}
The index is a finite value in the presence of spin-orbit interaction.
Thus, we conclude that $\check{H}_1$ satisfies necessary conditions (i) and (ii) 
in Sec.~\ref{ssec:chiral}. 
The zero-bias conductance in a dirty NS junction in Fig.~\ref{fig:system}(c) 
can be quantized as Eq.~(\ref{eq:gnsd}).
The shift of the Fermi surface by the spin-orbit interaction in 
Fig.~\ref{fig:fermi_surface} (a) makes the index be a nonzero value
because the integrand of Eq.~(\ref{eq:nzes_bd1}) is nonzero at (I) and (III). 
Therefore we will make clear what happen on the anomalous Green's function 
in these regions in Sec.~\ref{sec:f_function}.

\subsection{ Surface bound states of $\check{H}_2$}
\label{ssec:abs_h2}
Next, we focus on $\check{H}_2$ in Eq.~(\ref{eq:h2}).
The positive eigenvalues of $\check{H}_2$ are
calculated to be
\begin{align}
E_{2,\pm}= \sqrt{\xi_{2, \pm}^2 +\Delta_{\boldsymbol{k}}^2}, \quad
\xi_{2,\pm}(k, k_y) =& \xi_{\boldsymbol{k}}\pm \lambda\, k.
\end{align}
In Fig.~\ref{fig:fermi_surface}(b), we illustrate the two splitting Fermi surfaces
characterized by $\xi_{2,\pm}=0$.
The wave numbers in the $x$ direction on the Fermi surface are calculated as
\begin{align}
p_{2\pm} = \sqrt{k_F^2-k_y^2} \mp \alpha k_F.
\end{align} 
For $|k_y| \leq k_F$, the transport channels are propagating.
A topologically protected ZES appears for each propagating channel. 
The wave functions of such surface bound states are calculated as
\begin{align}
\psi_{ 2 +}(x)=&\left[\begin{array}{c} s_k\\ -is_k\\ 1 \\ -i \end{array}\right] 
e^{-q_{x}x}\sin p_{2+} x,\\\
\psi_{2 -}(x)=&\left[\begin{array}{c} s_k\\ is_k\\ -1 \\ -i \end{array}\right]
e^{-q_{x} x}\sin p_{2-} x. 
\end{align}
The Hamiltonian $\check{H}_2$ anticommutes to an extra chiral operator $\hat{\tau}_1$, 
which is derived from the invariance of $\check{H}_2$ under the rotation about the second 
axis in spin space $\hat{\sigma}_2\, \hat{\tau}_0$.
It is easy to show that $\hat{\tau}_1\, \psi_{ 2 \pm}
= \pm s_k \, \phi_{2 \pm}$.
We find $\mathcal{N}_{\mathrm{ZES}}=0$ because the chiral eigenvalues depend on $\mathrm{sgn}(k_y)$. 
Namely, $\check{H}_2$ does not satisfy necessary condition 
for the anomalous proximity effect (ii) in Sec.~\ref{ssec:chiral}. 
Thus we conclude the a superconductor described by $\check{H}_2$ does not exhibit 
the anomalous proximity effect because
the random impurity potential at its surface lifts the degeneracy 
at the zero energy.

\section{Pairing correlations at a surface}
\label{sec:f_function}
The purpose of this section is to check a close relationship 
between conditions (ii) and (iii) discussed in Sec.~\ref{sec:ape}.
As $\check{H}_1$ satisfies condition (ii), an odd-frequency spin-triplet
$s$-wave pairing correlation is expected to appear at a surface of a superconductor.
We will make clear mechanisms of generating such an odd-frequency pair.
As $\check{H}_2$ does not satisfy (ii), 
we will confirm that an odd-frequency spin-triplet
$s$-wave pairing correlation is absent at its surface.

To this end, we calculate the Green's function at a surface of a superconductor 
as shown in Fig.~\ref{fig:system}(b). 
The outline of the derivation is as follows.
We first solve the Gor'kov equation for the retarded Green's function,
\begin{align}
&\left[\epsilon+i\delta - \check{H}_{\mathrm{BdG}}(\boldsymbol{k}) \right]
\, \check{G}_{\epsilon}(\boldsymbol{k})
= \hat{\tau}_0\, \hat{\sigma}_0, \label{eq:gorkov}\\
&\check{G}_{\epsilon}(\boldsymbol{k})=\left[\begin{array}{cc}
\hat{G}_\epsilon(\boldsymbol{k}) & \hat{F}_\epsilon(\boldsymbol{k}) \\
-\undertilde{\hat{F}_\epsilon}(\boldsymbol{k}) &- \undertilde{\hat{G}_\epsilon}(\boldsymbol{k})
\end{array}\right]
\end{align}
where $i \delta$ is a small imaginary part of energy.
The 'undertilded' function is defined by 
\begin{align}
\undertilde{X_\epsilon}(\boldsymbol{k})\equiv {X}^\ast_{-\epsilon}(-\boldsymbol{k}),
\end{align}
where $X$ is an arbitrary function. 
Secondly, we calculate the Green's function in real space,
\begin{align}
\check{G}(\boldsymbol{r}, \boldsymbol{r}^\prime) =
\int_{-\infty}^{\infty}\frac{d k}{2\pi} e^{ik(x-x^\prime)} \sum_{k_y} \frac{e^{ik_y(y-y^\prime)}}{W} \,\check{G}(\boldsymbol{k}).
\label{eq:fourier}
\end{align}
for an infinitely long superconductor in the $x$ direction as shown in Fig.~\ref{fig:system}(a).
Finally, we introduce a high potential barrier 
$V \delta(x)\, \hat{\tau}_3$ to divide the infinitely long superconductor 
into two semi-infinitely long superconductors. 
One of them is illustrated in Fig.~\ref{fig:system}(b). 
The Green's function at a surface $\check{G}^{\mathrm{S}}(\boldsymbol{r}, \boldsymbol{r}^\prime)$ can be 
calculated by solving the Lippmann-Schwinger equation exactly 
as explained in Appendix~\ref{seca:ls_equation}.
The Green's function at a semi-infinite superconductor $\check{\mathbb{G}}$ is represented as
\begin{align}
\check{\mathbb{G}}(\boldsymbol{r}, \boldsymbol{r}^\prime) =
\check{G}(\boldsymbol{r}, \boldsymbol{r}^\prime) + \check{G}^{\mathrm{S}}(\boldsymbol{r}, \boldsymbol{r}^\prime).
\label{eq:g_surface}
\end{align}
In the text, we will show only the results of calculation 
and details of derivation are given in Appendix~\ref{asec:derivation}.

\subsection{ pairing correlations of $\check{H}_1$}
\label{ssec:pc_h1}

\begin{figure}[htbp]
\begin{center}
  \includegraphics[width=8.0cm]{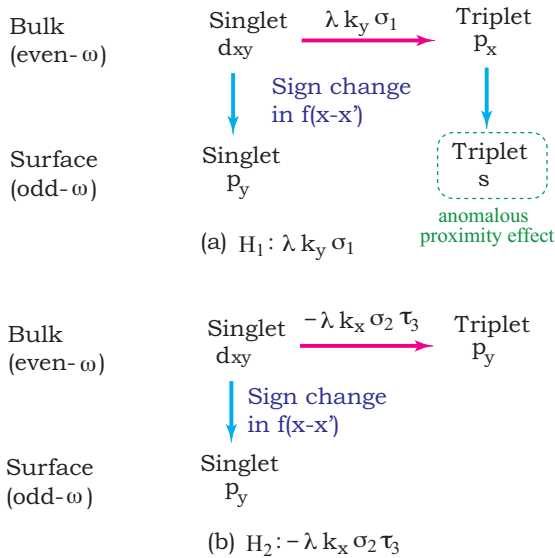}
\caption{The pairing correlations appearing in a $d_{xy}$-wave superconductor.
At a surface of $d_{xy}$-wave superconductor, an odd-frequency spin-singlet $p_y$-wave pairing correlation
always appear as a result of the sign change in the $d_{xy}$-wave 
pairing correlation under $x \leftrightarrow x^\prime$.
 In (a), a spin-orbit interaction $ \lambda k_y \hat{\sigma}_1$ induces spin-triplet $p_x$-wave 
 pairing correlation in bulk. An odd-frequency spin-triplet $s$-wave 
 component appears at a surface as a result of the sign change of the $p_x$-wave pairing correlation.
In (b), $-\lambda k_x \hat{\sigma}_2 \hat{\tau}_3$ induces spin-triplet $p_y$-wave 
 pairing correlation in bulk
 while this interaction does not generate 
 an odd-frequency spin-triplet $s$-wave correlation at a surface.
}
\label{fig:pairlist}
\end{center}
\end{figure}

We first study the pairing correlations at a surface of a superconductor
described by $\check{H}_1$.
The resulting anomalous Green's function at a surface is given by,  
\begin{widetext}
\begin{align}
\hat{F}^{\mathrm{S}}_{i\omega_n}(\boldsymbol{r}, \boldsymbol{r}^\prime)
=& \frac{1}{W}\sum_{k_y}  \, \frac{e^{ik_y(y-y^\prime)}}{2 \, v_F} e^{-q_x(x+x^\prime)} 
\left[
 \frac{i}{\Omega_n } \, \sin p_x(x-x^\prime) \, \Delta \, \bar{k}_y \, n_+(k_y) 
%
-\, \frac{i}{\Omega_n } \sin p_x(x-x^\prime) \, 
\Delta \, \bar{k}_y \, n_-(k_y) \, \hat{\sigma}_1 \right.\nonumber\\
& + \frac{2}{\omega_n} \sin p_x x\, \sin p_x x^\prime \, 
\Delta \, \bar{k}_y \, n_+(k_y)  
%
\left. - \frac{2}{\omega_n} \sin p_x x\, \sin p_x x^\prime \, 
\Delta \, \bar{k}_y\,  n_-(k_y) \, \hat{\sigma}_1 
\right]\, i \hat{\sigma}_2, \label{eq:f1_surface}
\end{align}
\end{widetext}
where $\Omega_n=\sqrt{\omega_n^2+\Delta_{\boldsymbol{k}}^2}$,
$p_x= \sqrt{k_F^2 - k_y^2 + 2 \alpha |k_y| k_F}$, and we applied the analytic continuation 
$\epsilon + i \delta \to i \omega_n$ to analyze the symmetry of pairing correlations.
We use the standard expression in Eq.~(\ref{eq:def_deco}) 
to decompose the anomalous Green's function into four spin components.
The first term in Eq.~(\ref{eq:f1_surface}) represents spin-singlet $d_{xy}$-wave 
pairing correlation because 
it changes sign under 
$x\leftrightarrow x^\prime$ and $y\leftrightarrow y^\prime$ independently. 
Such pairing correlation is linked to the pair potential in the presence of 
attractive interactions between two electrons.
The third term represents the pairing correlation
belongs to an odd-frequency spin-singlet $p_y$-wave symmetry class.
This correlation function changes the sign 
under $y \leftrightarrow y^\prime$, whereas it preserves the sign under $x \leftrightarrow x^\prime$.
The spin-singlet $d_{xy}$-wave component in the anomalous Green's function 
is an odd function of $x-x^\prime$, 
which is responsible for two phenomena at a surface: 
the appearance of highly degenerate bound states at the zero energy 
and the appearance of an odd-frequency $p_y$-wave pairing correlation. 
The first and the third terms in Eq.~(\ref{eq:f1_surface}) exist even in the absence of 
spin-orbit interactions.
The spin-orbit interaction in $\check{H}_1$ generates a spin-triplet $p_x$-wave symmetry at the second term
from a spin-singlet $d_{xy}$-wave pairing correlation. 
It is easy to confirm that the second term remains unchanged 
under $y\leftrightarrow y^\prime$ and changes its sign under $x\leftrightarrow x^\prime$.
In addition, the third term corresponds to an equal spin pairing 
correlation in the standard expression in Eq.~(\ref{eq:def_deco}).
The anomalous Green's function in an infinitely long superconductor consists of two pairing correlations.
One belongs to a spin-singlet $d_{xy}$-wave symmetry. The other belongs to a 
spin-triplet $p_x$-wave symmetry as shown in Eq.~(\ref{eq:f01m}).  
Since the spin-triplet $p_x$-wave pairing correlation is also an odd function of $x-x^\prime$, 
it induces a subdominant pairing correlation at a surface as shown in the last 
term of Eq.~(\ref{eq:f1_surface}). 
The relation among the four components are illustrated schematically in Fig.~\ref{fig:pairlist}(a).
The last pairing correlation in Eq.~(\ref{eq:f1_surface}) belongs to an odd-frequency 
spin-triplet $s$-wave symmetry.
It is easy to check that this component does not change its sign in 
$x \leftrightarrow x^\prime$ and $y \leftrightarrow y^\prime$ independently. 
As shown in $n_-(k_y)$ in Fig.~\ref{fig:index}(b), this odd-frequency correlation is nonzero 
at region (I) and (III) in $k_y$. 
This property is closely related to the fact that 
$g_{+}+g_{-}$ in Eq.~(\ref{eq:nzes_bd1}) is a nonzero value of -1 
at these regions as shown in Fig.~\ref{fig:index}(e).  
We conclude that spin-orbit interaction in $\check{H}_1$ generates an odd-frequency spin-triplet $s$-wave pairing correlation at a surface. 
Thus a superconductor described by $\check{H}_1$ satisfies also the last necessary condition (iii) for the anomalous 
proximity effect.

The local density of states at a surface 
is calculated from the normal Green's function as
\begin{align}
N_{\mathrm{S}}&(x,\epsilon) \equiv -\frac{1}{2\pi} \mathrm{Im}
\int_{-W/2}^{W/2} dy \mathrm{Tr}
\left[ \check{G}_{\epsilon}^{\mathrm{S}}(\boldsymbol{r}, \boldsymbol{r})\right].
\end{align}
The calculated results of the normal Green's function at a surface is given by
\begin{align}
\hat{G}^{\mathrm{S}}_{\epsilon}&(\boldsymbol{r},\boldsymbol{r}^\prime)= 
\sum_{k_y}
\frac{i\, e^{ik_y(y-y^\prime)} e^{-q_x(x+x^\prime)}}{2 W  v_{p_x} \Omega  }
\left[\epsilon \cos p_x (x+x^\prime)  \right.\nonumber\\
&\left.+i \, \Omega  \,\sin p_x(x+x^\prime)+ 2
\frac{\Delta^2_{\hat{\boldsymbol{k}}}}{\epsilon+i\delta} \,\sin p_xx \, \sin p_xx^\prime
 \right]\nonumber\\
 &\times \left[n_+(k_y) + n_-(k_y) \, \hat{\sigma}_1 \right]. \label{eq:gs_h1}
\end{align}
The LDOS at a surface results in
\begin{align}
N_{\mathrm{S}}(x,\epsilon) 
=&\delta(\epsilon)\sum_{k_y} \frac{2 |
\Delta_{{\boldsymbol{k}}}|}{ v_{p_x}} \left\{ \sin(p_x x) e^{-q_x x} \right\}^2 \nonumber\\
& \times n_+(k_y), \label{eq:ldos1}
\end{align}
 for $ \epsilon \ll \Delta$.
The local density of states at a surface has a peak at the zero energy, which reflects 
the presence of highly degenerate surface bound states at the zero energy.
The appearance of an odd-frequency pairing correlation and that of surface bound states 
at the zero energy are linked to each other directly. 
Mathematically, $\epsilon+i\delta$ in the denominator of $\hat{G}^{\mathrm{S}}$ in Eq.~(\ref{eq:gs_h1}) 
and $\omega_n$ in the denominator of the odd-frequency components of $\hat{F}^{\mathrm{S}}$ in Eq.~(\ref{eq:f1_surface})
 have the same origin within the analytic continuation $\epsilon+i\delta \to i\omega_n$. 
The transformation of 
\begin{align}
\frac{1}{\epsilon+i\delta}= \frac{P}{\epsilon}-i\pi \delta(\epsilon),
\end{align}
implies that the zero-energy peak in LDOS is a consequence of appearing 
an odd-frequency Cooper pair at a surface.

\subsection{Pairing correlations of $\check{H}_2$ }

%
%


The Green's function of a superconductor described by $\check{H}_2$ 
is also calculated by solving the Lippmann-Schwinger 
equation.
The anomalous Green's function at a surface results in
\begin{align}
F^{\mathrm{S}}_{i\omega_n}&(\boldsymbol{r}, \boldsymbol{r}^\prime)
= \frac{1}{W}\sum_{k_y} e^{i k_y (y-y^\prime)}
 \frac{e^{- q_x (x+x^\prime)}}{v_x}  
 \nonumber\\
&\times
\left[
i\, \frac{\Delta_{{\boldsymbol{k}}} }{\Omega_n } \sin k_x(x-x^\prime) + 
\frac{2 }{ \omega_n}
\Delta_{{\boldsymbol{k}}} \, \sin k_x x \, \sin k_x x^\prime 
\right.\nonumber\\
&
+\left.
\frac{\alpha \, \Delta\, \bar{k}_y }{\Omega_n } \cos k_x(x-x^\prime)\, \hat{\sigma}_2
\right] i\,  \hat{\sigma}_2.\label{eq:fsm2}
\end{align}
The first term belongs to spin-singlet $d_{xy}$-wave symmetry and is linked to the
pair potential. 
The second term is induced at a surface and belongs to odd-frequency spin-singlet $p_y$-wave symmetry.
Due to breaking local inversion symmetry at a surface, spin-singlet $p_y$-wave component 
is generated from spin-singlet $d_{xy}$-wave pairing correlation.
As already discussed in Eq.~(\ref{eq:f1_surface}), these two correlations exist in the absence of 
spin-orbit interactions.
The spin-orbit interaction in $\check{H}_2$ generates
a spin-triplet $p_y$-wave symmetry pairing correlation at the third term which 
changes sign under $y \leftrightarrow  y^\prime$ but retains its sign under $x \leftrightarrow x^\prime$.
The relations among the pairing correlations are illustrated in Fig.~\ref{fig:pairlist}(b). 
In Eq~(\ref{eq:fsm2}), however, an odd-frequency spin-triplet $s$-wave correlation is absent as we 
 expected at the beginning of this section.
The usual proximity effect is also absent in a normal metal 
attached to a superconductor because all of the pairing correlations in Eq.~(\ref{eq:fsm2}) are
odd functions of $y-y^\prime$~\cite{asano:prb2001-dwave,asano:jpsj2002,tanaka:prl2003}. 
Thus the spin-orbit interaction $\lambda k_x \hat{\sigma}_2\, \hat{\tau}_3$ does not contribute to 
the proximity effect.

\section{Discussion}
We have concluded that a superconductor described by $\check{H}_1$ in Eq.~(\ref{eq:h1}) 
satisfies all the necessary conditions for the anomalous proximity effect.
On the other hand, a superconductor described by $\check{H}_2$ in Eq.~(\ref{eq:h2}) 
does not exhibit any type of proximity effect.
Here we discuss briefly the proximity effect of a $d_{xy}$-wave superconductor
where the two spin-orbit interaction terms coexist and
form the Rashba spin-orbit interaction $\lambda k_y \hat{\sigma}_1 - \lambda k_x \hat{\sigma}_2\, \hat{\tau}_3$. 
Unfortunately, the coexistence of the two interaction terms 
weakens the anomalous proximity effect seriously because the interaction 
$- \lambda k_x \hat{\sigma}_2 \, \hat{\tau}_3$ breaks the extra chiral symmetry of $\check{H}_1$.
It is easy to confirm that $ - \lambda k_x \hat{\sigma}_2 \hat{\tau}_3$ does not 
anticommute to $\Gamma_{\mathrm{BDI}}$.
Therefore, random impurity potential lifts the degeneracy at the zero energy.
This explains why the signal of the anomalous proximity effect in Ref.~\onlinecite{tamura:prb2019} is 
very weak. Simultaneously, this discussion will determine the design guidelines for superconductors.
We conclude that a $d_{xy}$-wave superconductor described by $\check{H}_1$ is necessary
to observe the strong anomalous proximity effect such as the conductance quantization 
in a NS junction in Eq.~(\ref{eq:gnsd}) and the fractional current-phase relationship 
of the Josephson current in an SNS junction.

The control of spin-orbit interactions has been an important issue also in spintronics. 
The spin-orbit interaction in $\check{H}_1$ can be realized by tuning the 
Rashba spin-orbit interaction and the Dresselhaus spin-orbit interaction.
It has been known that such type of interaction 
stabilizes a specialized spin configuration in momentum space called persistent spin helix.
\cite{bernevig:prl2006,koralek:nature2009,kohda:prb2012,walser:natphys2012,schliemann:rmp2017}
Thus it would be possible to fabricate
a superconductor exhibits the strong anomalous proximity effect by combining existing 
technologies.

The anomalous proximity effect has been considered as a phenomenon unique to spin-triplet 
superconductors such as a $p_x$-wave superconductor in Eq.~(\ref{aeq:hpx}), 
a Majorana nanowire~\cite{sato:prb2006,sato:prb2009,lutchyn:prl2010,oreg:prl2010}, 
and two-dimensional artificial spin-triplet superconductors hosting flat-Majorana band at 
its surface~\cite{alicea:prb2010,you:prb2013}. 
It is no doubt that the existence of spin-triplet order parameter (pair potential) is a 
sufficient condition for a superconductor exhibiting the anomalous proximity effect.
On the other hand, we conclude that the necessary conditions discussed in Sec.~\ref{sec:ape} 
can be satisfied by a specific spin-triplet pairing correlation on a spin-singlet superconductor.
Our results indicate a way of enriching the superconducting properties of high-$T_c$ cuprate 
superconductors.

\section{Conclusion }
We theoretically studied the effects of spin-orbit 
interactions in a spin-singlet $d_{xy}$-wave superconductor on 
 symmetry of the pairing correlations and those on chiral properties of 
 the zero-energy bound states at a surface. 
The Hamiltonian for a $d_{xy}$-wave superconductor with a specific 
spin-orbit interaction preserves a chiral symmetry.
The chiral property of the surface bound states is analyzed in terms of an index 
$\mathcal{N}_{\mathrm{ZES}}$ which 
is a measure of the strength of the anomalous proximity effect.
Our results show that the spin-orbit interaction modifies the chiral property of 
the surface bound states drastically.
As a result, $\mathcal{N}_{\mathrm{ZES}}$ can be nontrivial nonzero values.
The symmetry of the pairing correlations at a surface of a superconductor are analyzed 
by using the anomalous Green's function which is obtained by solving the Gor'kov equation 
and the Lippmann-Schwinger equation analytically.
The spin-orbit interactions induce spatially uniform spin-triplet $p$-wave pairing correlations
in a $d_{xy}$-wave superconductor. 
A $p$-wave pairing correlation generates an odd-frequency
spin-triplet $s$-wave Cooper pair at a surface of a $d_{xy}$-wave superconductor.
We conclude that the presence of a spin-triplet pairing correlation 
causes the anomalous proximity effect even when a spin-triplet order parameter is absent.

To assist these conclusions, we should study low-energy transport properties 
through a dirty normal metal such as the conductance in a NS junction and 
the Josephson current in a SNS junction. 
The investigation by using numerical simulation is under way.
Results will be presented in somewhere else.

\begin{acknowledgments}
The authors are grateful to S.~Tamura, Y.~Tanaka and D.~Manske for useful discussion.
This work was supported by JSPS KAKENHI
(No.~JP20H01857), JSPS Core-to-Core Program (A.
Advanced Research Networks), and JSPS and Russian
Foundation for Basic Research under Japan-Russia
Research Cooperative Program Grant No. 19-52-50026.
\end{acknowledgments}

\appendix
\begin{widetext}
\section{Surface Andreev bound states}
\label{asec:sabs}
The BdG Hamiltonian on a two-dimensional tight-binding lattice
is represented by
\begin{align}
\mathcal{H}= \sum_{\boldsymbol{r}, \boldsymbol{r}^\prime}
\left[ \psi^\dagger_{\boldsymbol{r}, \uparrow},   \psi^\dagger_{\boldsymbol{r}, \downarrow}, 
\psi_{\boldsymbol{r}, \uparrow}, \psi_{\boldsymbol{r}, \downarrow} \right]
\check{H}_{\mathrm{BdG}}(\boldsymbol{r}, \boldsymbol{r}^\prime)
\left[ \psi_{\boldsymbol{r}^\prime, \uparrow},   \psi_{\boldsymbol{r}^\prime, \downarrow}, 
\psi^\dagger_{\boldsymbol{r}^\prime, \uparrow}, \psi^\dagger_{\boldsymbol{r}^\prime, \downarrow} \right]^{\mathrm{T}}
\end{align}
where $\psi_{\boldsymbol{r}, \sigma}$ is the annihilation operator of an electron 
at $\boldsymbol{r}=j \hat{\boldsymbol{x}} + m \hat{\boldsymbol{y}}$ with
with $\hat{\boldsymbol{x}}$ ($\hat{\boldsymbol{y}}$) being unit vector 
in the $\boldsymbol{x}$ ($\boldsymbol{y}$) direction, $\sigma=\uparrow$ or $\downarrow$ indicates spin of an electron, 
and $\mathrm{T}$ represents the transpose of a matrix.
The BdG Hamiltonian for an equal spin-triplet $p_x$-wave superconductor can be represented as
\begin{align}
\check{H}_{p_x}(\boldsymbol{r}, \boldsymbol{r}^\prime)=&{H}_{t}(\boldsymbol{r}, \boldsymbol{r}^\prime) \hat{\tau}_3 + 
\hat{H}_{\Delta_p}(\boldsymbol{r}, \boldsymbol{r}^\prime) \, \hat{\sigma}_3\, \hat{\tau}_1, \label{aeq:hpx}\\
{H}_t(\boldsymbol{r}, \boldsymbol{r}^\prime) = &
-t \left[
\delta_{\boldsymbol{r}, \boldsymbol{r}^\prime+\hat{\boldsymbol{x}}} + \delta_{\boldsymbol{r}, \boldsymbol{r}^\prime-\hat{\boldsymbol{x}}} 
+ \delta_{\boldsymbol{r}, \boldsymbol{r}^\prime+\hat{\boldsymbol{y}}} +\delta_{\boldsymbol{r}, \boldsymbol{r}^\prime-\hat{\boldsymbol{y}}} 
 \right] \hat{\sigma}_0
+ \delta_{\boldsymbol{r}, \boldsymbol{r}^\prime}  \left( 4t- \epsilon_F \right) \hat{\sigma}_0,\\
{H}_{\Delta_p}=& \frac{\Delta}{2i} 
 \left[ \delta_{\boldsymbol{r}, \boldsymbol{r}^\prime+\hat{\boldsymbol{x}}}  - \delta_{\boldsymbol{r}, \boldsymbol{r}^\prime-\hat{\boldsymbol{x}}} 
\right],
%
\end{align}
where $\tau_j$ and $\sigma_j$ for $j=1-3$ are the Pauli's matrices in particle-hole space and those in spin space, respectively.
The unit matrix in these spaces are $\hat{\tau}_0$ and $\hat{\sigma}_0$. 
The pair potential and the hopping integral are denoted by $\Delta$ and $t$, respectively.
We calculate the eigenvalues of $\hat{H}_{p_x}$ under 
the hard wall boundary condition in the $x$ direction and the periodic boundary
condition in the $y$ direction. 
The results for $\epsilon_F=2t$ and $\Delta=t$ are plotted 
as a function of $k_y$ in Fig.~\ref{fig:sabs}(a). 
The symbols for $E \neq 0$ represent eigenvalues of bulk states.
At the present parameter choice, the superconducting gap has two nodes at $k_y=\pm k_F$ with $k_F=\pi/2$.
The ZESs between the nodes in Fig.~\ref{fig:sabs}(a) localize at a surface.
The wave function of the ZESs near $x =0$ are described by
\begin{align}
\phi_{p_x, -}(\boldsymbol{r})=& 
\left[\begin{array}{c} 1 \\ 1 \\ i  \\ -i 
\end{array}\right] f_{k_y}(\boldsymbol{r}), \quad
\phi_{p_x, +}(\boldsymbol{r})= 
\left[\begin{array}{c} -1 \\ 1 \\ i \\ i 
\end{array}\right] f_{k_y}(\boldsymbol{r}),\quad
f_{k_y}(\boldsymbol{r}) = A\, e^{ik_y y} e^{-x/\xi_0}\, \sin k_x x, \label{aeq:phi_pm}
\end{align}
where $\xi_0 = v_F/ \Delta$ and $A$ is a constant.
It is easy to confirm that $\check{\Gamma}_{\mathrm{BIII}}  \,\phi_{\pm} = \pm \phi_\pm$. 
As a result, we find $N_+=N_-$ and $\mathcal{N}_{\mathrm{ZES}}=0$ in consistent 
with the prediction.~\cite{ikegaya:prb2018}

The wave functions of the ZESs in Eq.~(\ref{aeq:phi_pm}) are described alternatively as
\begin{align}
\phi_{p_x, \uparrow}(\boldsymbol{r})=& 
\left[\begin{array}{c} 1 \\ 0 \\ i  \\ 0 
\end{array}\right] f_{k_y}(\boldsymbol{r}), \quad
\phi_{p_x, \downarrow}(\boldsymbol{r})= 
\left[\begin{array}{c} 0 \\ 1 \\ 0 \\ -i 
\end{array}\right] f_{k_y}(\boldsymbol{r}), \label{aeq:phi_px}
\end{align}
where $\phi_{p_x, \sigma}$ the the wave function in spin $\sigma$ sector. 
It is easy to confirm that $\check{\Gamma}_{\mathrm{BDI}} \, \phi_{\sigma} =  \phi_{\sigma}$ holds for both 
$\sigma=\uparrow$ and $\downarrow$.
Therefore we find $N_+=N_c$ and $N_-=0$, which results in $\mathcal{N}_{\mathrm{ZES}}=N_c$
as long as $\{ \check{H}_{p_x}, \check{\Gamma}_{\mathrm{BDI}} \}=0$.

The eigenvalues of the BdG Hamiltonian of a $d_{xy}$-wave superconductor, 
\begin{align}
\check{H}_{d_{xy}}(\boldsymbol{r}, \boldsymbol{r}^\prime)=&{H}_{t}(\boldsymbol{r}, \boldsymbol{r}^\prime) \hat{\tau}_3 
+ H_{\Delta_d}(\boldsymbol{r}, \boldsymbol{r}^\prime) \,\hat{\sigma}_2\, \hat{\tau}_2, \label{aeq:hdxy}\\
H_{\Delta_d}(\boldsymbol{r}, \boldsymbol{r}^\prime)=& \frac{\Delta}{2} 
\left[
\delta_{\boldsymbol{r}, \boldsymbol{r}^\prime+\hat{\boldsymbol{x}}+\hat{\boldsymbol{y}}} 
+
\delta_{\boldsymbol{r}, \boldsymbol{r}^\prime-\hat{\boldsymbol{x}}-\hat{\boldsymbol{y}}} 
-
\delta_{\boldsymbol{r}, \boldsymbol{r}^\prime+\hat{\boldsymbol{x}}-\hat{\boldsymbol{y}}} 
-
\delta_{\boldsymbol{r}, \boldsymbol{r}^\prime-\hat{\boldsymbol{x}}+\hat{\boldsymbol{y}}} 
\right],
\end{align}
are shown in Fig.~\ref{fig:sabs}(b). There are three nodal points at $k_y=0$ and $\pm k_F$. 
A surface Andreev bound state appears for each propagating channel on the Fermi surface.
The wave functions of the ZESs are described by
\begin{align}
\phi_{d_{xy}, +}(\boldsymbol{r})=& 
\left[\begin{array}{c} 1 \\ -1 \\ i \, s_k \\ i \, s_k 
\end{array}\right] f_{k_y}(\boldsymbol{r}), \quad
\phi_{d_{xy}, -}(\boldsymbol{r})= 
\left[\begin{array}{c} 1 \\ 1 \\ -i \, s_k \\ i\, s_k 
\end{array}\right] f_{k_y}(\boldsymbol{r}), \quad s_k = \mathrm{sgn}(k_y), \label{aeq:phi_dxy}
\end{align}
It is easy to confirm that $\check{\Gamma}_{\mathrm{BDI}} \, \phi_{d_{xy}, \pm } =  \pm s_k\, 
\phi_{d_{xy}, \pm }$ hold.
The chiral eigenvalues of ZESs depend on the sign of $k_y$ in a $d_{xy}$-wave superconductor.
As displayed in Fig.~\ref{fig:sabs}(b), the number of the ZESs at $k_y>0$ is equal to that at $k_y<0$. 
Although $\check{H}_{d_{xy}}$ anticommutes to $\check{\Gamma}_{\mathrm{BDI}}$, 
the index $\mathcal{N}_{\mathrm{ZES}}$ is always zero in the absence of spin-orbit interactions.
The Hamiltonian $\check{H}_{d_{xy}}$ anticommutes also to another chiral operator $\hat{\tau}_1$.
We, however, find that $\mathcal{N}_{\mathrm{ZES}}=0$ in terms of the chiral eigen values of $\hat{\tau}_{1}$.

\begin{figure}[tbp]
\begin{center}
  \includegraphics[width=7.0cm]{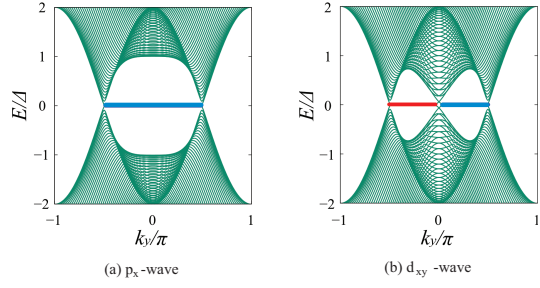}
\caption{
The eigenvalues of the BdG Hamiltonian on a two-dimensional tight-binding model 
are plotted as a function of the wave number in the $y$ direction.
The results for a $p_x$-wave superconductor in (a) have nodes at $k_y = \pm k_F$ with $k_F= \pi/2$. 
The surface Andreev bound states are degenerate at $E=0$ as indicated by a line 
between the two nodal points. All of the ZESs belong to the same chiral eigenvalue of $\check{\Gamma}_{\mathrm{BDI}}$.
In (b), the results for a $d_{xy}$-wave superconductor are shown.
An  additional nodal point appears at $k_y=0$. 
ZESs for $k_y>0$ and those 
for $k_y<0$ belong to the opposite chiral eigenvalue of $\check{\Gamma}_{\mathrm{BDI}}$. 
 }
\label{fig:sabs}
\end{center}
\end{figure}

\section{Green's function in real space}
\label{asec:derivation}
The solution of the Gor'kov equation in Eq.~(\ref{eq:gorkov}) is obtained as
\begin{align}
\hat{G}_\epsilon(\boldsymbol{k})=&\left[ 
(\epsilon - \hat{H}_{\mathrm{N}}) + \hat{\Delta}\, 
(\epsilon + \undertilde{\hat{H}}_{\mathrm{N}})^{-1}\, \undertilde{\hat{\Delta}} 
\right]^{-1}, \quad
\hat{F}_\epsilon(\boldsymbol{k})=\left[ 
\undertilde{\hat{\Delta}} + 
(\epsilon + \undertilde{\hat{H}}_{\mathrm{N}})
\,  \hat{\Delta}^{-1}\, 
(\epsilon - \hat{H}_{\mathrm{N}})
\right]^{-1}.
\end{align}
The retarded Green's function for $H_1$ is represented as
\begin{align}
\hat{G}_{\epsilon}(\boldsymbol{r}-\boldsymbol{r}^\prime) = &
\frac{1}{W}\sum_{k_y}\, e^{i k_y (y-y^\prime)} \hat{G}_{\epsilon}(x-x^\prime),\quad
\hat{F}_{\epsilon}(\boldsymbol{r}-\boldsymbol{r}^\prime) = 
\frac{1}{W}\sum_{k_y}\, e^{i k_y (y-y^\prime)} \hat{F}_{\epsilon}(x-x^\prime),
\end{align}
with
\begin{align}
\hat{G}_{\epsilon}&(x-x^\prime) =
\frac{1}{2\pi} \int_{-\infty}^\infty dk\, e^{ik(x-x^\prime)} \frac{1}{2}
\left[ \frac{(\epsilon+\xi_{1,+})(1+\hat{\sigma}_1) }{(\epsilon+i\delta)^2-E_{1,+}^2} 
+ \frac{(\epsilon+\xi_{1, -})(1-\hat{\sigma}_1) }{(\epsilon+i\delta)^2-E_{1, -}^2} 
\right],\label{eq:gk1e}\\
\hat{F}_{\epsilon}&(x-x^\prime) 
= \frac{1}{2\pi} \int_{-\infty}^\infty dk\, e^{ik(x-x^\prime)}\frac{\Delta_{{\boldsymbol{k}}}}{2} 
\left[ \frac{1 -\hat{\sigma}_1}{(\epsilon+i\delta)^2 - E_{1, -}^2} 
+ \frac{1 +\hat{\sigma}_1}{(\epsilon+i\delta)^2 - E_{1, +}^2}
\right] \, i\hat{\sigma}_2. \label{eq:fk1e}
\end{align}
To proceed the calculation, it is necessary to carry out the integration 
\begin{align}
I_{1,\pm}= \frac{1}{2\pi} \int_{-\infty}^\infty dk\, e^{ik(x-x^\prime)}\, 
\frac{f(k, \xi_{1,\pm})}{(\epsilon +i \delta  - E_{1,\pm})(\epsilon+E_{1,\pm})}, \label{eq:ipm_def}
\end{align}
where $f$ is an analytic function. 
For $\epsilon>0$, $(\epsilon  + E_{1,\pm})^{-1}$ is analytic, whereas 
$(\epsilon +i \delta  - E_{1,\pm})^{-1}$ has two poles at $k=k_\pm^e+i0^+$ and $k=-k_\pm^h+i0^+$ with
\begin{align}
k_\pm^e=& \left[ p_{1\pm}^2
+ \frac{2m}{\hbar^2} \Omega \right]^{1/2}, \quad
k_\pm^h= \left[ p_{1\pm}^2
- \frac{2m}{\hbar^2} \Omega\right]^{1/2}, 
\quad \Omega=\sqrt{(\epsilon+i\delta)^2 - \Delta^2_{\boldsymbol{k}} }
\end{align}
where $p_{1\pm}$ is defined in Eq.~(\ref{eq:p1_def}). 
Paying attention to the relations
$\xi_{1,\pm} \to \Omega  $ at $k=k^e_\pm$ 
and $\xi_{1,\pm} \to -\Omega$ at $k=-k^h_\pm$,  
$\epsilon +i \delta  - E_{1,\pm}$ at the denominator of Eq.~(\ref{eq:ipm_def}) can be expanded around 
$k=k^e_\pm$ as
\begin{align}
\epsilon + i \delta - E_{1,\pm}(k) 
\approx \epsilon - E_{1,\pm}(k_\pm^e) + i\delta - \frac{\xi_{1,\pm}}{E_{1,\pm}(k_\pm^e)}
\frac{ \hbar^2 k_\pm^e}{m}(k - k_\pm^e)
=  - \frac{\Omega}{\epsilon} \frac{ \hbar^2 k_\pm^e}{m} \,(k - k_\pm^e - i \delta^\prime).   
\end{align}
Around a pole of $k=-k_\pm^h $, the $\epsilon +i \delta  - E_{1,\pm}$ becomes
\begin{align}
\epsilon + i \delta - E_{1,\pm}(k) 
\approx \epsilon  - E_{1,\pm}(-k_\pm^h) + i\delta  - 
\frac{\xi_{1,\pm}}{E_{1,\pm}(-k_\pm^h)}\frac{ -\hbar^2 k_\pm^h}{m} (k+k_\pm^h)
=  - \frac{-\Omega}{\epsilon} \frac{(-\hbar^2 k_\pm^h)}{m} \, (k+k_\pm^h -i\delta^\prime).   
\end{align}
By picking up the residues of these poles,
the integral in Eq.~(\ref{eq:ipm_def}) is calculated as
\begin{align}
I_{1,\pm} =& -i \frac{m}{2\Omega \hbar^2 } 
\left[
\frac{ f( s_x k_{\pm}^e, \Omega)}{k_{\pm}^e} e^{ik_\pm^e |x-x^\prime|} 
+ \frac{f( - s_x k_{\pm}^h, -\Omega)}{k_{\pm}^e} e^{-ik_\pm^h|x-x^\prime|} 
\right]\Theta(p_{1\pm}^2),\\
s_x=&\mathrm{sgn}(x-x^\prime).
\end{align}
where we consider the contribution from the propagating channels as 
indicated by a factor $\Theta(p_{1\pm}^2)$. 
The integral on the upper-half complex plane converges for $x-x^\prime>0$.
On the other hand for $x-x^\prime<0$, the transformation of $k \to -k$ is necessary, which
produces a factor of $s_x$. 
In the text, we mainly discuss the Green's function for $0 < \epsilon \ll \Delta$.
In such case, it is possible to apply an approximation for the wavenumber
$k^e_\pm = p_{1\pm} + i q_{x}$ and $k_\pm^h= p_{1\pm}-i q_{x}$ with $q_x=(\Delta/\hbar v_F) \bar{k}_y$.
We also apply $p_{1+} \approx p_{1-} \approx p_x = \sqrt{k_F^2-k_y^2 + 2\alpha k_F |k_y|}$ 
because the amplitude of the wavenumber is not important for analyzing the pairing symmetry.
By the same reason, we consider $q_x$ only in the exponential function. 
The integral under these approximations is represented as
\begin{align}
I_{1,\pm} \approx & -i \frac{m}{2\Omega \hbar^2 } e^{-q_x |x-x^\prime|}
\left[
\frac{ f( s_x p_{x}, \Omega)}{p_{x}} e^{i p_x |x-x^\prime|} 
+ \frac{f( - s_x p_x, -\Omega)}{p_{x}} e^{-i p_x|x-x^\prime|} 
\right]\Theta(p_{1\pm}^2).
\end{align}

The normal Greens's function in Eq.~(\ref{eq:gk1e})
is calculated as
\begin{align}
G_\epsilon(x-x^\prime)=&
-i \frac{m}{4\Omega \hbar^2 }  e^{-q_x |x-x^\prime|}
\left[
\frac{\epsilon + \Omega}{p_x} e^{i p_x |x-x^\prime|} 
+\frac{\epsilon - \Omega}{p_x} e^{-i p_x |x-x^\prime|} 
\right](1+\hat{\sigma}_1) \Theta(p_{1+}^2)\nonumber \\
&-i \frac{m}{4\Omega \hbar^2 }  e^{-q_x |x-x^\prime|}
\left[
\frac{\epsilon + \Omega}{p_x} e^{i p_x |x-x^\prime|} 
+\frac{\epsilon - \Omega}{p_x} e^{-i p_x |x-x^\prime|} 
\right](1-\hat{\sigma}_1) \Theta(p_{1-}^2),\\
=&-i \frac{e^{-q_x |x-x^\prime|}}{2\Omega \hbar v_{p_x} }  
\left[ \epsilon \, \cos (p_x|x-x^\prime|) +i\, \Omega \, \sin (p_x|x-x^\prime|)\right]
\left[n_+(k_y) + n_-(k_y) \hat{\sigma}_1 \right].
\end{align}
The anomalous Green's function results in
\begin{align}
F_\epsilon(x-x^\prime)=&
-i \frac{m s_x}{4\Omega \hbar^2 }  e^{-q_x |x-x^\prime|}
\left[
\frac{\Delta \bar{k}_y \bar{p}_x }{p_x} e^{i p_x |x-x^\prime|} 
-
\frac{\Delta \bar{k}_y \bar{p}_x }{p_x} e^{-i p_x |x-x^\prime|} 
\right](1+\hat{\sigma}_1) \Theta(p_{1+}^2)i\hat{\sigma}_2\nonumber \\
&
-i \frac{m s_x}{4\Omega \hbar^2 }  e^{-q_x |x-x^\prime|}
\left[
\frac{\Delta \bar{k}_y \bar{p}_x }{p_x} e^{i p_x |x-x^\prime|} 
-
\frac{\Delta \bar{k}_y \bar{p}_x }{p_x} e^{-i p_x |x-x^\prime|} 
\right](1-\hat{\sigma}_1) \Theta(p_{1-}^2)\,i\hat{\sigma}_2 ,\\
=&-i \frac{e^{-q_x |x-x^\prime|}}{2\Omega \hbar v_{F} }  i \Delta \bar{k}_y
\sin p_x(x-x^\prime)
\left[n_+(k_y) + n_-(k_y) \hat{\sigma}_1 \right]i\hat{\sigma}_2.
\end{align}

To obtain the Green's function at a surface, the four parts of 
the Green's in a uniform superconductor are necessary.
We supply them as follows, 
\begin{align}
\hat{G}_{i\omega_n}(x-x^\prime)
= &- \frac{m}{2\Omega_n \hbar^2} 
\frac{e^{- q_{x} |x-x^\prime| }}{p_x} 
\left[ i\hbar\omega_n  \, \cos (p_x|x-x^\prime|) -\, \Omega_n \, \sin (p_x|x-x^\prime|) \right]
\left[n_+(k_y) +n_-(k_y) \hat{\sigma}_1 \right],\label{eq:g01m}\\
\hat{F}_{i\omega_n}(x-x^\prime)
= & - \frac{m }{2\Omega_n \hbar^2} 
\frac{ e^{-q_{x}|x-x^\prime|}}{p_x} \, 
\Delta \bar{p}_x \bar{k}_y \,i \sin p_x(x-x^\prime) 
\left[ n_+(k_y) + n_-(k_y)\,  \hat{\sigma}_1
\right]i\, \hat{\sigma}_2, \label{eq:f01m}\\
-\undertilde{\hat{G}}_{i\omega_n}(x-x^\prime)
= &- \frac{m}{2\Omega_n \hbar^2} 
\frac{e^{- q_{x} |x-x^\prime| }}{p_x}
\left[ i\hbar\omega_n  \, \cos (p_x|x-x^\prime|) +\, \Omega_n \, \sin (p_x|x-x^\prime|) \right]
\left[n_+(k_y) - n_-(k_y) \hat{\sigma}_1 \right],\label{eq:g01mt}\\
-\undertilde{\hat{F}}_{i\omega_n}(x-x^\prime)
= & - \frac{m }{2\Omega_n \hbar^2} 
\frac{ e^{-q_{x}|x-x^\prime|}}{p_x} \, 
\Delta \bar{p}_x \bar{k}_y \,i \sin p_x(x-x^\prime) 
\left[ - n_+(k_y) + n_-(k_y)\,  \hat{\sigma}_1
\right]i\, \hat{\sigma}_2,\label{eq:f01mt}
\end{align}
where $\Omega_n=\sqrt{(\hbar\omega_n)^2+\Delta_{\boldsymbol{k}}^2}$ and the particle-hole transformation is expressed by
\begin{align}
\undertilde{X}_{i\omega_n}(x, k_y) =X^\ast_{i\omega_n}(x, -k_y), 
\end{align}
in this representation.

The solution of the Gor'kov equation for $\check{H}_2$ is represented as
\begin{align}
\hat{G}_{\epsilon}&(x-x^\prime) =
\frac{1}{2\pi} \int_{-\infty}^\infty dk\, e^{ik(x-x^\prime)} \frac{1}{2}
\left[ \frac{(\epsilon+\xi_{2,+})(1+\hat{\sigma}_2) }{(\epsilon+i\delta)^2-E_{2,+}^2} 
+ \frac{(\epsilon+\xi_{2, -})(1-\hat{\sigma}_2) }{(\epsilon+i\delta)^2-E_{2, -}^2} 
\right],\\
\hat{F}_{\epsilon}&(x-x^\prime) 
= \frac{1}{2\pi} \int_{-\infty}^\infty dk\, e^{ik(x-x^\prime)}
\frac{\Delta_{{\boldsymbol{k}}}}{2} 
\left[ \frac{1 -\hat{\sigma}_2}{(\epsilon+i\delta)^2 - E_{2, -}^2} 
+ \frac{1 +\hat{\sigma}_2}{(\epsilon+i\delta)^2 - E_{2, +}^2}
\right] \, i\hat{\sigma}_2.
\end{align}
Within the first order $\alpha$, 
the retarded Green's functions in an infinitely long superconductor 
are calculated as,
\begin{align}
\hat{G}_{\epsilon}&(x-x^\prime)= \frac{-i}{\Omega \, \hbar v_{k_x}} e^{-q_x|x-x^\prime| }
\left[
\epsilon \cos k_x|x-x^\prime| +i  \Omega \sin k_x|x-x^\prime|
\right],\\
\hat{F}_{\epsilon}&(x-x^\prime)= \frac{-i}{\Omega\, \hbar  v_{k_x}} e^{-q_x|x-x^\prime| }
\left[
i \Delta_{{\boldsymbol{k}}} \sin k_x(x-x^\prime)
+\alpha \, \Delta\, \bar{k}_y 
\cos k_x (x-x^\prime) \hat{\sigma}_2 
\right]\, i\hat{\sigma}_2.
\end{align}
Here we apply the relation,
\begin{align}
I_{2,\pm} =&\frac{1}{2\pi} \int_{-\infty}^\infty dk\, e^{ik(x-x^\prime)}\, 
\frac{f(k, \xi_{2,\pm})}{(\epsilon +i \delta  - E_{2,\pm})(\epsilon+E_{2,\pm})},\\
=& -i \frac{e^{-q_x |x-x^\prime|}}{2\Omega \hbar v_x } 
\left[
 f( \mp \alpha k_F+ s_x k_x, \Omega)\,  e^{ik_x|x-x^\prime|} 
+ f( \mp \alpha k_F- s_x k_x, -\Omega)\, e^{-ik_x|x-x^\prime|} 
\right],
\end{align}
where $v_x=\hbar k_x/m$ and $k_x=\sqrt{k_F^2-k_y^2}$ is a real wavenumber.

\section{Lippmann-Schwinger equation}
\label{seca:ls_equation}
The Lippmann-Schwinger equation relates the Green's function in the presence of a 
perturbation $\check{\mathbb{G}}$
to that in the absence of the perturbation $\check{G}$ as
\begin{align}
\check{\mathbb{G}}_{i\omega_n}(\boldsymbol{r}, \boldsymbol{r}^\prime) = &
\check{G}_{i\omega_n}(\boldsymbol{r}, \boldsymbol{r}^\prime)
+\int d \boldsymbol{r}_1\, \check{G}_{i\omega_n}(\boldsymbol{r}, \boldsymbol{r}_1) 
\check{V}(\boldsymbol{r}_1) \, 
\check{\mathbb{G}}_{i\omega_n}(\boldsymbol{r}_1, \boldsymbol{r}^\prime),
\end{align}
where $\check{V}(\boldsymbol{r})$ is the perturbation potential.
In this paper, we introduce a wall at $x=x_0$ 
\begin{align}
\check{V}(\boldsymbol{r}) = V\, \delta(x-x_0)\, \hat{\tau}_3,
\end{align}
to divide an infinitely long superconductor into two semi-infinite superconductors.
Thus $G_{i\omega_n}$ is the Green's function in a infinitely long superconductor in the $x$ direction.
Although the wall breaks the translational symmetry in the $x$ direction, the superconductor is 
translational invariant in the $y$ direction. Therefore it is possible to represent the Green's function 
as
\begin{align}
\check{\mathbb{G}}_{i\omega_n}(\boldsymbol{r}, \boldsymbol{r}^\prime) = \frac{1}{W} \sum_{k_y} 
\check{\mathbb{G}}_{i\omega_n}(x, x^\prime)
e^{ik_y(y-y^\prime)}. \label{eq:gky}
\end{align}
By substituting the expression into the equation, we find
\begin{align}
\check{\mathbb{G}}_{i\omega_n}(x, x^\prime) &= \check{G}_{i\omega_n}(x, x^\prime)
+\check{G}_{i\omega_n}(x, x_0)\, V\hat{\tau}_3\, \check{\mathbb{G}}_{i\omega_n}(x_0, x^\prime).
\end{align}
By putting $x=x_0$, we obtain the Green's function,
\begin{align}
\check{\mathbb{G}}_{i\omega_n}(x_0, x^\prime) &= 
\left[ 1- \check{G}_{i\omega_n}(x_0, x_0) \, V\, \hat{\tau}_3 \right]^{-1}
\check{G}_{i\omega_n}(x_0, x^\prime).
\end{align}
Finally, we reach an relation,
\begin{align}
\check{\mathbb{G}}_{i\omega_n}(x, x^\prime) &= \check{G}_{i\omega_n}(x, x^\prime) 
+ \check{G}_{i\omega_n}^{\mathrm{S}}(x, x^\prime), \\
\check{G}_{i\omega_n}^{\mathrm{S}}(x, x^\prime)&= \check{G}_{i\omega_n}(x, x_0)\, 
V\hat{\tau}_3\, 
\left[ 1- \check{G}_{i\omega_n}^{(0)}(x_0, x_0) \, V\, \hat{\tau}_3 \right]^{-1}\, 
 \check{G}_{i\omega_n}(x_0, x^\prime).\label{eq:gs_def}
\end{align}
The Green's function at a surface of a superconductor 
is calculated by taking a limit of $V \to \infty$.
In the text, we input $x_0=0$ and analyze Eq.~(\ref{eq:gs_def}) near the surface 
$0< x \lesssim \xi_0$ and $0< x^\prime \lesssim \xi_0$ with $\xi_0=\hbar v_F/ \pi \Delta$ 
being the coherence length.
\end{widetext}

%


\end{document}